\documentstyle[11pt,epsfig]{article}
\def\edth{\;\raise1.0pt\hbox{$'$}\hskip-6pt\partial\;}
\def\baredth{\;\overline{\raise1.0pt\hbox{$'$}\hskip-6pt
\partial}\;}
\def\gsim{~\rlap{$>$}{\lower 1.0ex\hbox{$\sim$}}}

\def\be{\begin{equation}}
\def\ee{\end{equation}}
\def\ba{\begin{eqnarray}}
\def\ea{\end{eqnarray}}

\def\sk{\rm{\sk}}
\newcommand{\fr}[2]{\frac{#1}{#2}}

\begin{document}

\title{Growth index with the cosmological constant}

\author{Seokcheon Lee$^{\,1,2}$ and Kin-Wang Ng$^{\,1,2,3}$}

\maketitle

$^1${\it Institute of Physics, Academia Sinica,
Taipei, Taiwan 11529, R.O.C.}

$^2${\it Leung Center for Cosmology and Particle Astrophysics, National Taiwan University, Taipei, Taiwan 10617, R.O.C.}

$^3${\it Institute of Astronomy and Astrophysics,
 Academia Sinica, Taipei, Taiwan 11529, R.O.C.}

\begin{abstract}
We obtain the exact analytic form of the growth index at present
epoch ($a=1$) in a flat universe with the cosmological constant
({\it i.e.} the dark energy with its equation of state
$\omega_{de} = -1$). For the cosmological constant, we obtain the
exact value of the current growth index parameter $\gamma =
0.5547$, which is very close to the well known value $6/11$. We
also obtain the exact analytic solution of the growth factor for
$\omega_{de}$ = $-1/3$ or $-1$. We investigate the growth index
and its parameter at any epoch with this exact solution. In
addition to this, we are able to find the exact $\Omega_{m}^{0}$
dependence of those observable quantities. The growth index is quite sensitive to $\Omega_{m}^{0}$ at $z = 0.15$, where we are able to use 2dF observation. If we adopt 2dF value of growth index, then we obtain the constrain $0.11 \leq \Omega_{m}^{0} \leq 0.37$ for the cosmological constant model. Especially, the growth index is quite sensitive to $\Omega_{m}^{0}$ around $z \leq 1$. We might be able to obtain interesting observations around this epoch. Thus, the analytic solution for this growth factor provides the very useful tools for
future observations to constrain the exact values of observational
quantities at any epoch related to growth factor for $\omega_{de}
= -1$ or $-1/3$.

\end{abstract}

The background evolution equations in a flat
Friedmann-Robertson-Walker universe ($\rho_m + \rho_{de} =
\rho_{cr}$) are \ba \Bigl(\fr{\dot{a}}{a}\Bigr)^2 &=& \fr{8\pi
G}{3}(\rho_{m} + \rho_{de}) = \fr{8 \pi G}{3} \rho_{cr} \, ,
\label{H} \\ 2 \fr{\ddot{a}}{a} + \Bigl(\fr{\dot{a}}{a}\Bigr)^2
&=& - 8 \pi G \omega_{de} \rho_{de} \, , \label{dotH} \ea where
$\omega_{de}$ is the equation of state (eos) of dark energy,
$\rho_{m}$ and $\rho_{de}$ are the energy densities of the matter
and the dark energy, respectively. We consider the constant
$\omega_{de}$ only. The dark energy does not participate directly
in cluster formation, but it alters the cosmic evolution of the
background. The linear density perturbation of the matter ($\delta
= \delta \rho_{m} / \rho_{m}$) for sub-horizon scales is governed
by \cite{Bonnor} \be \ddot{\delta} + 2 \fr{\dot{a}}{a}
\dot{\delta} = 4 \pi G \rho_{m} \delta \, . \label{ddotdelta} \ee
The textbook solution for the growing mode solution of Eq.
(\ref{ddotdelta}) for $\omega_{de} = -1$ or $-1/3$ is \cite{Heath,
Varun, Dodelson} \be \delta_{g} (a) = \fr{5 \Omega_{m}^{0}}{2}
\fr{H(a)}{H_{0}} \int_{0}^{a} X^{-3/2}(a') da' \, , \label{deltag}
\ee where $X(a) = (aH/H_{0})^2 = \Omega_{m}^{0} a^{-1} +
\Omega_{de}^{0} a^{-1-3\omega_{de}}$, $\Omega_{m}^{0} =
\rho_{m}^{0}/\rho_{cr}^{0}$ and $\Omega_{de}^{0} = 1 -
\Omega_{m}^{0}$ are the present energy density contrast of the
matter and the dark energy, respectively. The growth index $f$ is
defined as \be f = \fr{d \ln \delta_{g}}{d \ln a} \equiv
\Omega_{m}(a)^{\gamma} \, . \label{f} \ee Thus, $f$ is expressed
from Eq. (\ref{deltag}) as \be f(\Omega_{m}^{0}, \omega_{de}, a) =
-\fr{3}{2} - \fr{3 \omega_{de} (1 - \Omega_{m}^{0})
a^{-1-3\omega_{de}}}{2X} + \fr{a X^{-3/2}(a)}{\int_{0}^{a}
X^{-3/2}(a') da'} \, . \label{f2} \ee We obtain the exact analytic
form of the growth index for $\omega_{de} = -1$ or $-1/3$ at the
present epoch ($a=1, \, X(a=1)=1$) as \be f(\Omega_{m}^{0},
\omega_{de}, 1) = -\fr{3}{2} - \fr{3}{2} \omega_{de} (1 -
\Omega_{m}^{0}) - 3 \omega_{de} (\Omega_{m}^{0})^{\fr{3}{2}}
\fr{\Gamma[1-\fr{5}{6\omega_{de}}] /
(\Gamma[\fr{-5}{6\omega_{de}}]\Gamma[1])}{F[\fr{3}{2},
\fr{-5}{6\omega_{de}}, 1-\fr{5}{6\omega_{de}},
-\fr{\Omega_{de}^0}{\Omega_{m}^{0}}]} \, , \label{f0} \ee where
$\Gamma$ is the gamma function and $F$ is the hypergeometric
function. We emphasize that this formula is correct only for
$\omega_{de} = -1/3$ or $-1$. For $\Omega_{m}^{0} =1$ one get $f =
1$ for all a, which is consistent with $\delta \propto a$. For
$\omega_{de} = -\fr{1}{3}$, $f$ is matched with a non-flat without
the cosmological constant $f(a=1) = -1 - \Omega_{m}^{0}/2 + 5/2
(\Omega_{m}^{0})^{3/2}/F[3/2,5/2,7/2,1-(\Omega_{m}^{0})^{-1}]$ as
shown in Ref.~\cite{Lahav}. For the cosmological constant, the
above formula becomes \be f(\Omega_{m}^{0}, -1, 1) \equiv
f_{L}^{0} = -\fr{3}{2} \Omega_{m}^{0} + 3
(\Omega_{m}^{0})^{\fr{3}{2}} \fr{\Gamma[\fr{11}{6}] /
\Gamma[\fr{5}{6}]}{F[\fr{3}{2}, \fr{5}{6}, \fr{11}{6},
-\fr{\Omega_{de}^0}{\Omega_{m}^{0}}]} \, , \label{f0L} \ee where
we use $\Gamma[1] =1$. From the exact form of the growth index
$f_{L}^{0}$, we are able to obtain the exact present value of the
growth index parameter $\gamma_{L}^{0}$, \be \gamma_{L}^{0} \equiv
\fr{\ln f_{L}^{0}}{\ln \Omega_{m}^{0}} = \fr{\ln \Bigl[ -\fr{3}{2}
\Omega_{m}^{0} + (\Omega_{m}^{0})^{\fr{3}{2}}
\fr{\Gamma[\fr{11}{6}] / \Gamma[\fr{5}{6}]}{F[\fr{3}{2},
\fr{5}{6}, \fr{11}{6}, -\Omega_{de}^0/ \Omega_{m}^{0}]} \Bigr]}{
\ln \Omega_{m}^{0}} \, . \label{gamma0} \ee
\begin{center}
\begin{figure}
\vspace{1.5cm} \centerline{ \psfig{file=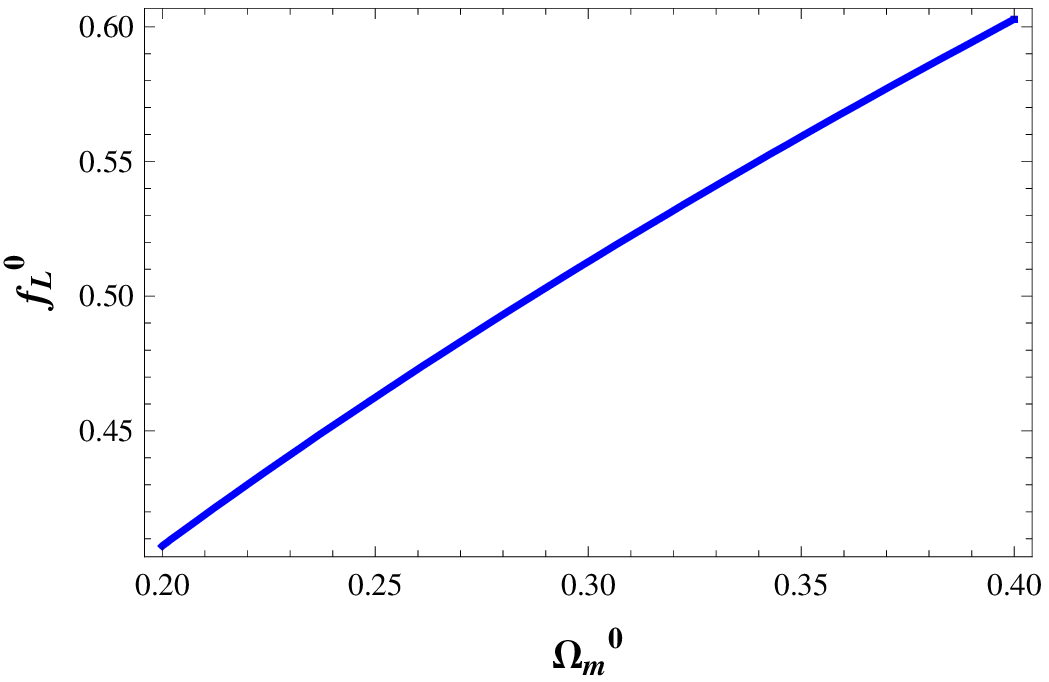, width=6.5cm}
\psfig{file=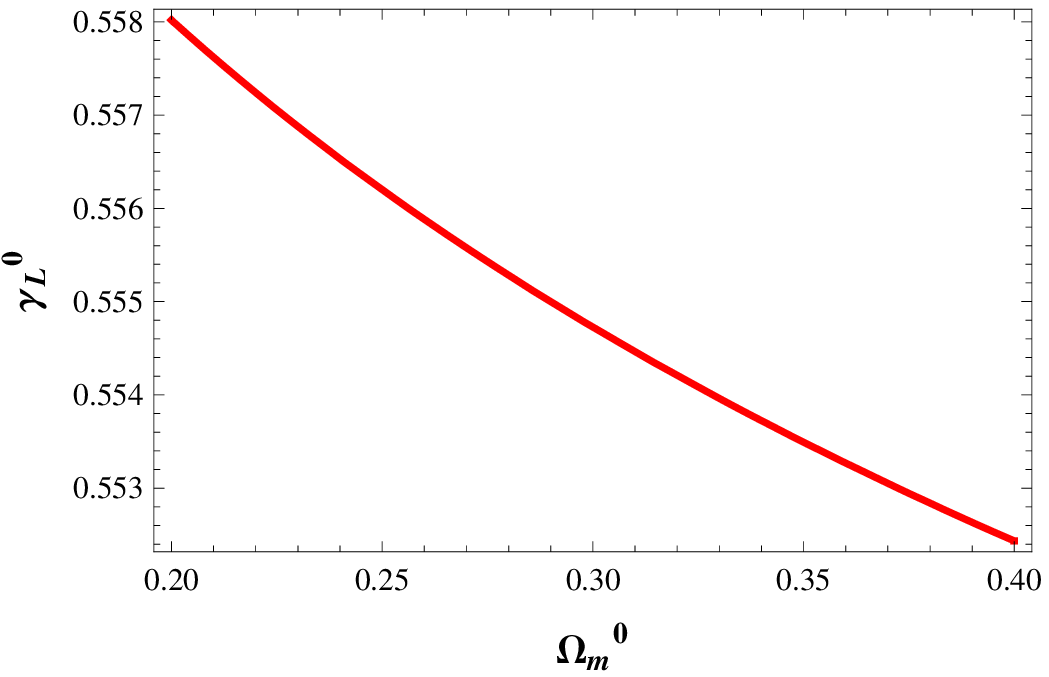, width=6.5cm} } \vspace{-0.1cm} \caption{
$\Omega_{m}^{0}$ dependence of $f_{L}^{0}$ and $\gamma_{L}^{0}$.
a) Current value of the growth index $f_{L}^{0}$ for $0.2 \leq
\Omega_{m}^{0} \leq 0.4$. b) $\gamma_{L}^{0}$ for the same range
of $\Omega_{m}^{0}$. } \label{fig1}
\end{figure}
\end{center}
In Fig. \ref{fig1}, we show the present value of the growth index
$f_{L}^{0}$ and its parameter $\gamma_{L}^{0}$ as a function of
$\Omega_{m}^{0}$. In Fig.~\ref{fig1}a, we show the variation of
the present growth index for different values of $\Omega_{m}^{0}$.
It is changed from $0.4073$ to $0.6028$ for $0.2 \leq
\Omega_{m}^{0} \leq 0.4$. Thus, there is up to $32 \%$ differences
in $f_{L}^{0}$ for the different values of $\Omega_{m}^{0}$.
However, this $\Omega_{m}^{0}$ dependence is decreased by using
the parameter $\gamma_{L}^{0}$. We show this variation of
$\gamma_{L}^{0}$ in Fig. \ref{fig1}b. $\gamma_{L}^{0}$ is changed
from $0.5580$ to $0.5524$ for $\Omega_{m}^{0} = 0.2$ and $0.4$,
respectively. We obtain $\gamma_{L}^{0} = 0.5547$ for
$\Omega_{m}^{0} = 0.3$.

We also obtain the exact analytic solution of the growth factor
given in Eq. (\ref{deltag}) for $\omega_{de}=-1/3,-1$ (see
Appendix for details). The solution of Eq.~(\ref{correction}) for
$\omega_{de} = -1$ is \be \delta_{g}^{L}(a) = c_{1}^{L} \sqrt{1 +
Q a^{-3}} + c_{2}^{L} Q^{\fr{2}{3}} a^{-2} F[1, \fr{1}{6},
\fr{5}{3}, - Q a^{-3}] \, , \label{deltaskDL2} \ee where
$c_{1,2}^{L}$ are integral constants and $Q = \Omega_{m}^{0}
/\Omega_{de}^{0}$. We need to determine the $c_{1}^{L}$ and
$c_{2}^{L}$ based on the proper initial conditions
$\delta_{g}^{L}(a_i) = a_{i}$ and $d \delta_{g}^{L} / da |_{a_i} =
1$ in the matter dominated epoch in order to make this solution
become the growing mode solution. We obtain from the initial
conditions that $c_{1}^{L} = 1.085464$ and $c_{2}^{L} = -0.943314$
when $\omega_{de} = -1.0$ and $\Omega_{m}^{0} = 0.3$.
\begin{center}
\begin{figure}
\vspace{1.5cm} \centerline{ \psfig{file=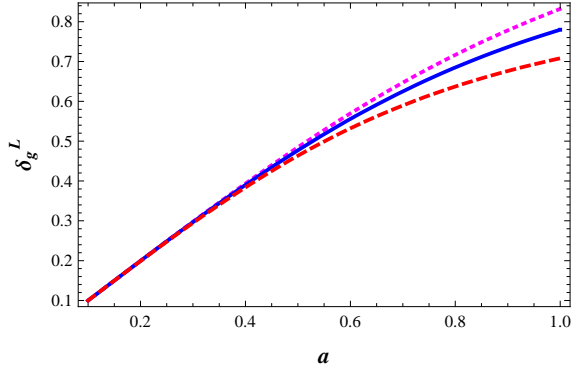, width=7.5cm} }
\vspace{-0.1cm} \caption{Cosmological evolution of
$\delta_{g}^{L}$ for $\Omega_{m}^{0}=0.4,0.3,0.2$ (from top to
bottom).} \label{fig2}
\end{figure}
\end{center}
We show the cosmological evolution of the growth factor
$\delta_{g}^{L}$ for different values of $\Omega_{m}^{0}$ in
Fig.~\ref{fig2}. The dotted, solid, and dashed lines correspond to
$\Omega_{m}^{0} = 0.4, 0.3$ and $0.2$ respectively. As
$\Omega_{m}^{0}$ increases, $\delta_{g}^{L}$ maintains the linear
growth with $a$ for a longer time, as expected. From the initial
conditions, we obtain $(c_{1}^{L}, c_{2}^{L}) = (1.25633,
-1.09263)$ and $(0.906875, -0.788711)$ for $\Omega_{m}^{0} = 0.4$
and $0.2$, respectively. We use these values of the coefficients
in Fig.~\ref{fig2}.

\begin{center}
\begin{figure}
\vspace{1.5cm} \centerline{ \psfig{file=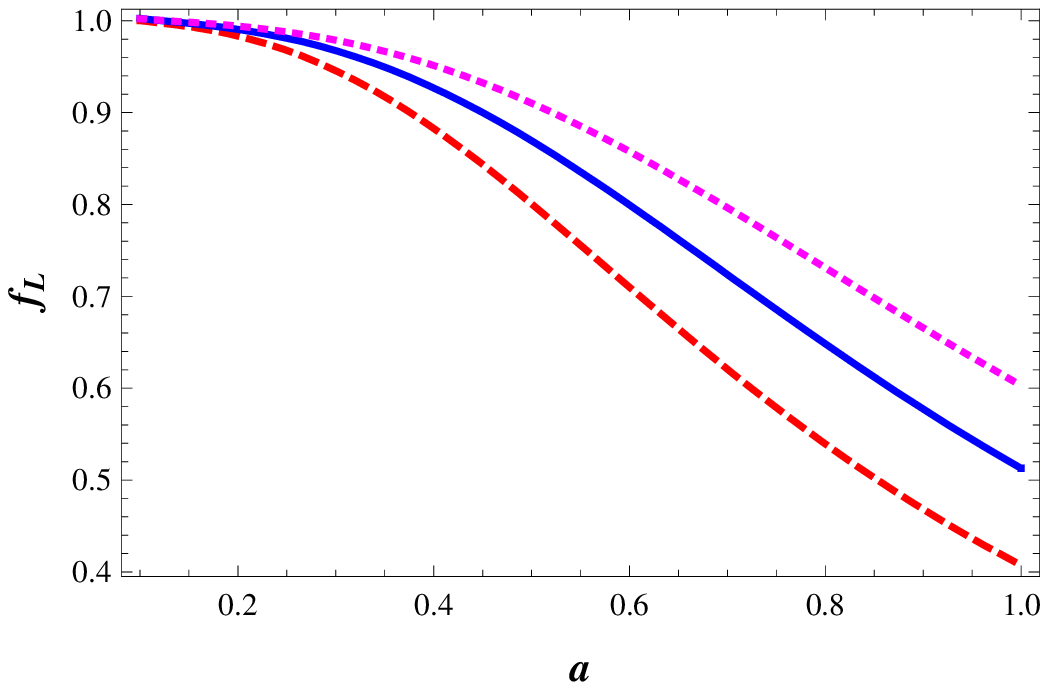,
width=6.5cm}\psfig{file=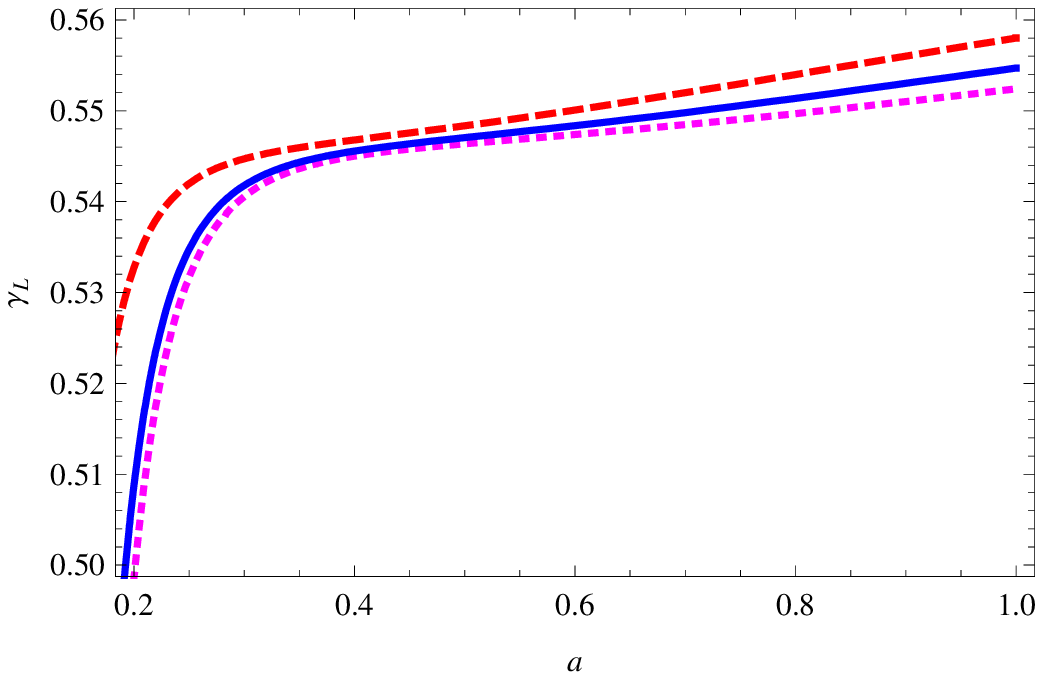, width=6.5cm} } \vspace{-0.1cm}
\caption{ a) Evolution of $f_{L}(a)$ for different values of
$\Omega_{m}^{0}$. b) The same for $\gamma_{L}(a)$. Dotted, solid,
and dashed lines correspond to $\Omega_{m}^{0} = 0.4, 0.3$, and
$0.2$, respectively.} \label{fig3}
\end{figure}
\end{center}

\begin{center}
\begin{figure}
\vspace{1.5cm} \centerline{ \psfig{file=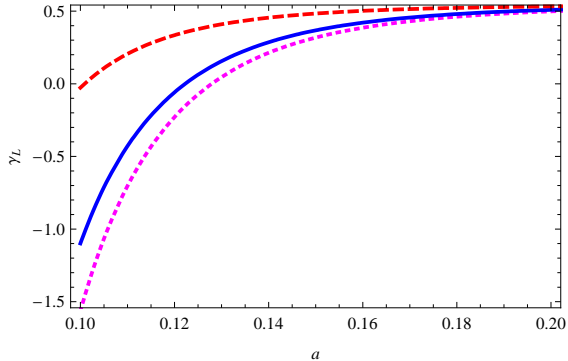, width=7.5cm} }
\vspace{-0.1cm} \caption{Cosmological evolution of
$\gamma_{L}(a)$ for $\Omega_{m}^{0}=0.2,0.3,0.4$ (from top to
bottom).} \label{fig4}
\end{figure}
\end{center}

In addition to the present value of the growth index $f(a=1)$, we
are able to obtain the growth index in any epoch from the exact
analytic solution of the growth factor~(\ref{deltaskDL2}). For
$\omega_{de} = -1$, we find that
\ba f_{L}(a) &=& \fr{d \ln \delta_{g}^{L}(a)}{d \ln a} = \fr{\Biggl( -\fr{3}{2} \fr{A[a]^2 - (c_{1}^{L})^2}{A[a]} - 2 B[a] + \fr{3}{10} Qa^{-3} B[a] \fr{F2}{F1} \Biggr)}{ (A[a]+B[a]) } \nonumber \\ \rm{where} \, , && A[a] = c_{1}^{L} \sqrt{1 + Q a^{-3}} \,\, , \,\,  B[a] = c_{2}^{L} Q^{\fr{3}{2}} a^{-2} F1 \label{faskL} \\
&& F1 = F[1, \fr{1}{6}, \fr{5}{3}, -Q a^{-3}] \,\, , \,\,  F2
=F[2, \fr{7}{6}, \fr{8}{3}, -Q a^{-3}] \nonumber \ea Even though
it looks complicated, one is able to obtain $f_{L}(a)$ by using
the solution $\delta_{g}^{L}$ without any difficulty. We are able
to find the growth index parameter $\gamma_{L}(a) = \fr{\ln
f_{L}(a)}{\ln \Omega_{m}(a)}$ in any epoch from this exact
analytic form of $f_{L}(a)$. We also investigate the
$\Omega_{m}^{0}$ dependence of them without any ambiguity. Thus,
this analytic solution is very useful for the investigation of
observational quantities. We show these properties in
Fig.~\ref{fig3}.
In Fig. \ref{fig3}a, we show the growth index $f_{L}(a)$ for
different values of $\Omega_{m}^{0}$. The dotted, solid and
dashed lines correspond to the evolution of $f_{L}(a)$ for
$\Omega_{m}^{0} = 0.4, 0.3$ and $0.2$, respectively. As we have
more matter at present, we have faster growing. Thus, we have
bigger values of $f_{L}$ when we have bigger values of
$\Omega_{m}^{0}$ as shown in Fig. \ref{fig3}a. The
$\Omega_{m}^{0}$ dependence is quite sensitive around $a \simeq
0.87$ ({\it i.e.} $z  \simeq 0.15$). Thus, the observed value of
$f(a=0.87) = 0.51 \pm 0.15$ from the 2dF galaxy redshift survey will be a good guideline
to measure $\Omega_{m}^{0}$ if the dark energy is the cosmological
constant \cite{2dF}. However, the present value of the growth index parameter
is insensitive to $\Omega_{m}^{0}$ as we also obtain in the
previous formula. The growth index parameter is sensitive to $\Omega_{m}^{0}$ around $a \geq 0.8$ ({\it i.e.} $z \leq 0.25$) as shown in Fig. \ref{fig3}b. Again, the dashed, solid and dotted lines correspond to the cosmological evolution of $\gamma_{L}(a)$ for $\Omega_{m}^{0} = 0.2, 0.3$ and $0.4$, respectively. Especially, the growth index parameter changes dramatically for $a \leq 0.3$ ({\it i.e.} $z \geq 2.3$). We show this in Fig. \ref{fig4}. The dashed, solid and dotted lines (from top to bottom) correspond to the evolution of $\gamma_{L}(a)$ for $\Omega_{m}^{0} = 0.2, 0.3$ and $0.4$, respectively.
We summarize the results in Table 1. As it is well known that the growth index parameter $\gamma_{L}$ is insensitive to $\Omega_{m}^{0}$ and $a$ up to $a > 0.3$ \cite{WS}. However, $\gamma_{L}(a)$ shows the strong model dependence around $a \leq 0.3$ ({\it i.e.} $z \geq 2.3$).  If we naively adopt the 2dF value of $f$ without considering the data error from selection effects, then we obtain the constrain $0.11 \leq \Omega_{m}^{0} \leq 0.37$ for the cosmological constant model.
\begin{center}
    \begin{table}
    \begin{tabular}{ | c | c | c | c | c |}
    \hline
    $\Omega_{m}^{0}$ & $f_{L}^{z=0}$ & $\gamma_{L}^{z=0}$ & $f_{L}^{z=0.15}$ & $\gamma_{L}^{z=0.15}$ \\ \hline
    0.2 & 0.407 & 0.558 & 0.488 & 0.555 \\ \hline
    0.3 & 0.513 & 0.555 & 0.598 & 0.553 \\ \hline
    0.4 & 0.603 & 0.552 & 0.685 & 0.551 \\ \hline
    \end{tabular}
    \caption{$\Omega_{m}^{0}$ is the present value of the matter density contrast, $f_{L}^{z=0}$ and $f_{L}^{z=0.15}$ correspond to the growth index at the present and $z = 0.15$, respectively. $\gamma_{L}^{z=0}$ and $\gamma_{L}^{z=0.15}$ are the growth index parameter at the present and $z=0.15$, respectively.}
    \label{table1}
    \end{table}
\end{center}
In linear theory, the peculiar velocity $\vec{v}_{pec}$ is related
to the peculiar acceleration $\vec{g}$ and/or interior average
over-density $\langle \delta\rangle_{R}$ in a spherical
perturbation of radius $R$, \cite{Peebles} \be |\vec{v}_{pec}| =
\fr{2}{3} \Bigl| \fr{f\vec{g}}{H\Omega_{m}^{0}} \Bigr| \, =
\fr{1}{3} HR f \langle \delta\rangle_{R} \, . \label{vpec} \ee The
difference of this peculiar velocity between two different values
of $\Omega_{m}^{0}$ is as large as $29$\% when we use the
different values of $f_{L}^{z=0.15}$ for $\Omega_{m}^{0} = 0.2$
and $0.4$. Thus, the exact analytic form of $f$ provides the good
analysis tool for galaxy redshift survey.

\appendix
\section{Appendix}

From Eq.~(\ref{f2}), we need to solve the integration to find the
analytic form of $f$ at the present epoch~\cite{Abramowitz}, \ba
\int_{0}^{1} \fr{da}{X^{3/2}(a)} &=& \int_{0}^{1}
\fr{da}{(\Omega_{m}^{0} a^{-1} + \Omega_{de}^{0}
a^{-1-3\omega_{de}})^{3/2}} = \int_{0}^{1} \fr{da}{(1 +
\Omega_{de}^{0}/\Omega_{m}^{0} a^{-3\omega_{de}})^{3/2}
(\Omega_{m}^0/a)^{3/2}} \nonumber \\ &=& \fr{-1}{3 \omega_{de}
(\Omega_{m}^{0})^{3/2}} \int_{0}^{1} \eta^{-1-5/(6 \omega_{de})}
(1 - r \eta)^{-3/2} d \eta \nonumber \\ &=& \fr{-1}{3 \omega_{de}
(\Omega_{m}^{0})^{3/2}} \fr{\Gamma[\fr{-5}{6\omega_{de}}]
\Gamma[1]}{\Gamma[1-\fr{5}{6 \omega_{de}}]} F[\fr{3}{2},
\fr{-5}{6\omega_{de}},
1-\fr{5}{6\omega_{de}},-\fr{\Omega_{de}^{0}}{\Omega_{m}^{0}}] \, ,
\label{int} \ea where we use $a^{-3\omega_{de}} = \eta$ and $r =
-\Omega_{de}^{0}/\Omega_{m}^{0} = 1 - (\Omega_{m}^{0})^{-1}$. Note
that Eq.~(\ref{int})is valid for any value of $\omega_{de}$. We
can also check the result of a non-flat universe without the
cosmological constant in Ref.~\cite{Lahav}. In this case we have
$k/H_{0}^2 \equiv \Omega_{K} = 1 - \Omega_{m}^{0}$. Then we obtain
the expression for $X = \Omega_{m}^{0} a^{-1} + \Omega_{K}$.
Mathematically, this is the same as the dark energy case with the
equation of state of the dark energy $\omega_{de} = -1/3$. Then
the above integral (\ref{int}) becomes \be \int_{0}^{1}
\fr{da}{X^{3/2}(a)} = \fr{1}{(\Omega_{m}^{0})^{3/2}}
\fr{\Gamma[\fr{5}{2}] \Gamma[1]}{\Gamma[\fr{7}{2}]} F[\fr{3}{2},
\fr{5}{2}, \fr{7}{2},-\fr{\Omega_{K}}{\Omega_{m}^{0}}] = \fr{2}{5}
(\Omega_{m}^{0})^{-3/2} F[\fr{3}{2}, \fr{5}{2}, \fr{7}{2},
1-(\Omega_{m}^{0})^{-1}] \, . \label{intK} \ee Thus, we can
reproduce $f = -1 -\Omega_{m}^{0}/2 +5/2
(\Omega_{m}^{0})^{3/2}/F[\fr{3}{2},\fr{5}{2},\fr{7}{2},1-(\Omega_{m}^{0})^{-1}]$
in a non-flat without the cosmological constant case.

It is better for us to rewrite the linear density perturbation
equation~(\ref{ddotdelta}) with changing the variable from $t$ to
$a$ to get \cite{Dodelson} \be \fr{d^2 \delta}{da^2} + \Biggl(
\fr{d \ln H}{d a} + \fr{3}{a} \Biggr) \fr{d \delta}{d a} - \fr{4
\pi G \rho_{m}}{(aH)^2} \delta = 0 \, . \label{dadelta} \ee When
we replace the growing mode solution (\ref{deltag}) into
Eq.~(\ref{ddotdelta}), we get \be \ddot{\delta}_{g}^{ex} + 2 H
\dot{\delta}_{g}^{ex} - 4 \pi G \rho_{m} \delta_{g}^{ex} - \Bigl[
4 \pi G (1 + \omega_{de})(1 + 3\omega_{de}) \rho_{de} \Bigr]
\delta_{g}^{ex} \, . \label{correction} \ee Note that we have
changed the notation from $\delta_{g}$ to $\delta_{g}^{ex}$ to
reflect the fact that $\delta_{g}^{ex}$ is the solution of
Eq.~(\ref{correction}) for any value of $\omega_{de}$. In
particular, when $\omega_{de} = -1$ or $-1/3$, the extended
solution $\delta_{g}^{ex}$ is reduced to $\delta_{g}$. After
replacing new parameters $Y = (\Omega_{m})/(\Omega_{de}) =
(\Omega_{m}^{0})/(\Omega_{de}^{0}) a^{3 \omega_{de}}$ in
Eq.~(\ref{correction}), we have \be Y \fr{d^2
\delta_{g}^{ex}}{dY^2} + \Bigl[1 + \fr{1}{6 \omega_{de}} -
\fr{1}{2(Y+1)} \Bigr] \fr{d \delta_{g}^{ex}}{dY} - \Bigl[\fr{1}{6
\omega_{de}^2 Y} + \fr{3\omega_{de} + 4}{6 \omega_{de} Y(Y+1)}
\Bigr] \delta_{g}^{ex} = 0 \, . \label{correction2} \ee Now we try
$\delta_{g}^{ex}(Y) = c_{1} \delta_{1}(Y) + c_{2} \delta_{2}(Y) =
c_{1} Y^{i} (1+Y)^{j} + c_{2} Y^{k} B(Y)$, where $c_{1,2}$ are
integral constants, because it is the most general combination of
the solution for the above equation (\ref{correction2}). We obtain
the constraints for $i$ and $j$ from $\delta_{1}$\, , \be i =
-\fr{1 + \omega_{de}}{2 \omega_{de}} \,\, , \hspace{0.2in} j =
\fr{1}{2} \, . \label{ij} \ee
We also replace $\delta_{2}$ into the above equation (\ref{correction2}) to get  \ba && Y (1 + Y) \fr{d^2B}{dY^2} + \Biggl[ \fr{5}{6 \omega_{de}} + \fr{5}{2} + \Bigl( 3 + \fr{5}{6 \omega_{de}} \Bigr) Y \Biggr] \fr{d B}{dY} + \Bigl( 1 + \fr{5}{6 \omega_{de}} \Bigr) B = 0 \,, \nonumber \\
&& {\rm when} \,\,\, \, k = 1 + \fr{1}{3 \omega_{de}} \,\, .
\label{correction3} \ea There are two alternative ways to make the
above equation as the hypergeometric differential equation, $Y =
-X$ or $1+Y = X$ \cite{Morse}. We choose the first case which
shows the proper behavior. Now the above equation
(\ref{correction3}) becomes the so-called hypergeometric
differential equation, \be (X)(1-X) \fr{d^2 B}{dX^2} + \Biggl[
\fr{1}{2} - \fr{18 \omega_{de} + 5}{6 \omega_{de}} X \Biggr] \fr{d
B}{d X} - \Bigl( 1 + \fr{5}{6 \omega_{de}} \Bigr) B = 0 \, ,
\label{hyper} \ee with the solution $B(X) = F[1, 1 + \fr{5}{6
\omega_{de}}, \fr{5}{2} + \fr{5}{6 \omega_{de}}, X]$ being the
hypergeometric function~\cite{Morse}. Thus, the full extended
solution becomes \be \delta_{g}^{ex}(Y) = c_{1}
Y^{-\fr{1+\omega_{de}}{2 \omega_{de}}} \sqrt{1 + Y} + c_{2}
Y^{\fr{1+3 \omega_{de}}{3 \omega_{de}}} F[1, 1 + \fr{5}{6
\omega_{de}}, \fr{5}{2} + \fr{5}{6 \omega_{de}}, - Y ] \, .
\label{deltask} \ee Writing $Y = a^{3 \omega_{de}} Q$, $Q =
\Omega_{m}^{0} /\Omega_{de}^{0}$, for $\omega_{de} = -1$ and
$-1/3$, we have \ba \delta_{g}^{ex}(a)_{\omega_{de}=-1} &=&
\delta_{g}^{L}(a)= c_{1}^{L} \sqrt{1 + Q a^{-3}} + c_{2}^{L}
Q^{\fr{2}{3}} a^{-2}
F[1, \fr{1}{6}, \fr{5}{3}, - Q a^{-3}] \, , \label{deltaskDL} \\
\delta_{g}^{ex}(a)_{\omega_{de}=-\fr{1}{3}} &=& c_{1} Q a^{-1}
\sqrt{1 + Q a^{-1} } + c_{2} F[1, -\fr{2}{3}, 0 , - Q a^{-1}] \, . \label{deltaskDw13}  \ea The
solution~(\ref{deltask}) is a mathematical one. We need to
further specify this solution in order to make it as a physical
solution. If we use the fact that growth factor $\delta \propto a$ in the matter
dominated epoch, then we can determine $c_{1}$ and $c_{2}$ to make
the solution as the growing mode solution. We are also able to
find the decaying mode solution if we use the decaying mode initial
conditions into this solution (\ref{deltask}).

\end{document}